\documentclass{article}
\usepackage{graphicx}
\usepackage{subcaption}
\usepackage{multicol}
\usepackage[left=1in,right=1in,top=1in,bottom=1in]{geometry}
\usepackage{cite}
\usepackage[colorlinks=true,urlcolor=blue,citecolor=blue,linkcolor=blue]{hyperref}
\usepackage[noabbrev,capitalise]{cleveref}

\title{\Large The Mirage of Breaking MIRAGE:\\ Refuting the HPCA'23 Paper ``Are Randomized Caches Truly Random?"}
\date{}

\author{
\begin{minipage}[t]{0.5\textwidth}
\centering
Gururaj Saileshwar \\
University of Toronto \& NVIDIA Research \\
\texttt{gururaj@cs.toronto.edu}
\end{minipage}
\hfill
\begin{minipage}[t]{0.5\textwidth}
\centering
Moinuddin Qureshi \\
Georgia Tech \\
\texttt{moin@gatech.edu}
\end{minipage}
}

\begin{document}

\maketitle

\begin{abstract}
The HPCA'23~\cite{randcachebroken} paper's claim that ``MIRAGE is broken'' relied on two faulty assumptions: (1)~starting from a severely compromised initial state where some sets are already full, and (2) a buggy cipher that does not provide uniformity of randomizing addresses over the cache sets. When we fixed these two shortcomings (starting with valid state and using AES/PRINCE cipher) we do not observe any conflict misses.  We provide the analysis below and share the code of our analysis along with the paper. 
\end{abstract}

\section{Description of MIRAGE}

MIRAGE~\cite{mirage} is a randomized cache that provides the abstraction of a fully associative randomized LLC with random-replacement. It eliminates set-conflicts, the root cause of conflict-based attacks, by provisioning invalid tags in the LLC and using load-balancing techniques inspired by Power of 2 Random Choices~\cite{power-of-2-choices} to ensure the availability of invalid tags. This allows new lines to be installed in invalid tags without set conflicts and evicted lines to be selected randomly in a fully associative manner, leaking no information. MIRAGE ensures set-conflicts are unlikely in the lifetime of the universe and eliminates conflict-based attacks.
\begin{figure}[h]
    \centering
    \vspace{-0.1in}
    \includegraphics[width=0.8\textwidth]{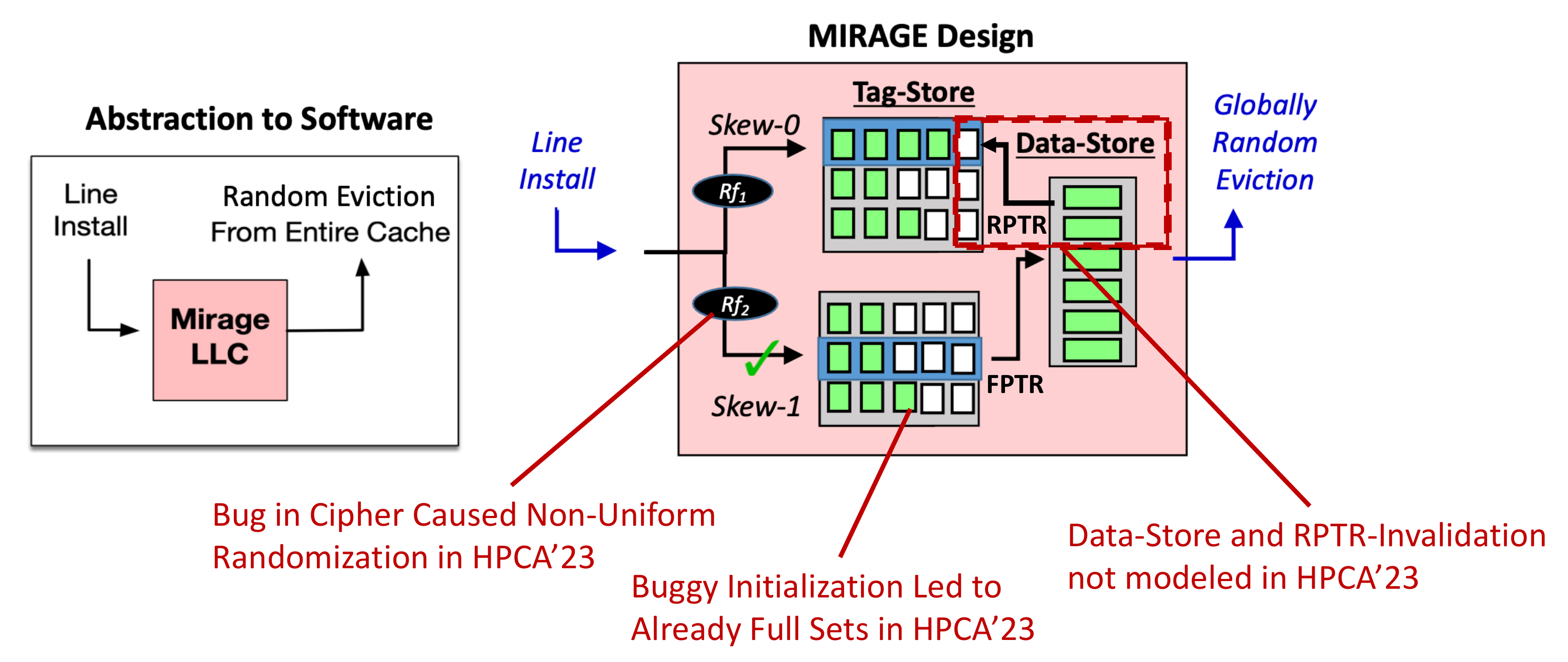}
    \vspace{-0.1in}
     \caption{Abstraction/Design of Mirage. Components in {\color{red} red text} were incorrectly modeled in HPCA'23~\cite{randcachebroken}.}
     \vspace{-0.1in}
    \label{fig:debunk_mirage}
\end{figure}

\section{Refuting claims of ``Are Randomized Caches Truly Random"}
In a recent paper titled ``Are Randomized Caches Truly Random? Formal Analysis of Randomized-Partitioned Caches,"~\cite{randcachebroken} published at HPCA'23, the authors claim to have ``broken" MIRAGE, by inducing set-associative evictions. However, on inspecting their code\footnote{\url{https://github.com/SEAL-IIT-KGP/randCache}}, we found fundamental bugs in their implementation that were the root-cause of these set-conflicts. 
Specifically, the results in Figure 6 and Figure 7 of their paper are impacted by these bugs.
Upon fixing these bugs, we were unable to observe any set-associative evictions, highlighting that MIRAGE is indeed not broken. Below, we describe the bugs we identified in the HPCA'23 paper's code and how we addressed them.

The first bug affecting the results of both Figure 6 and Figure 7 of the HPCA 2023 paper is the missing modeling of Global Evictions in MIRAGE, as shown in our ~\cref{fig:debunk_mirage}. This component was not modeled in either the bucket and balls model of the cache and the python cache simulator. Global evictions in MIRAGE are important as they not only randomly evict a data entry from the cache, but also the associated tag in the tag-store via a reverse-pointer (RPTR) lookup. This crucially reduces high occupancy sets and prevents an adversary from biasing the state towards high occupancy sets. Below, we discuss this and other bugs in each of the HPCA'23 experiments. 
All of our code is available at this link below\footnote{\url{https://github.com/gururaj-s/refuting_HPCA23_randCache}}.

\subsection{Bugs in the Bucket and Balls Simulation}

In our email to the authors of the HPCA'23 paper on February 14, 2023, we pointed out that their results from bucket and balls simulations in Figure 6 were incorrect as the simulator inserts more balls in buckets than the cache capacity. By adding an assert statement to check the number of balls in the buckets, we were able to detect this violation.  The root cause of this bug is the lack of modeling of global evictions in MIRAGE and consequently the lack of ball removal on each ball insertion. 
\cref{fig:bnb_assert_failure} shows the assert failure on running the Bucket and Balls simulation code from the paper's repository. 

In response, the authors provided an updated code with an  implementation of $remove\_ball()$ function. However, this implementation had a second bug in the array indexing logic that resulted in balls being removed from only a subset of the buckets, not all of the buckets. After we identified and fixed this bug, we were able to produce correct results.

\begin{figure}[htbp]
    \centering
    \begin{subfigure}[b]{0.49\textwidth}
        \includegraphics[width=\textwidth]{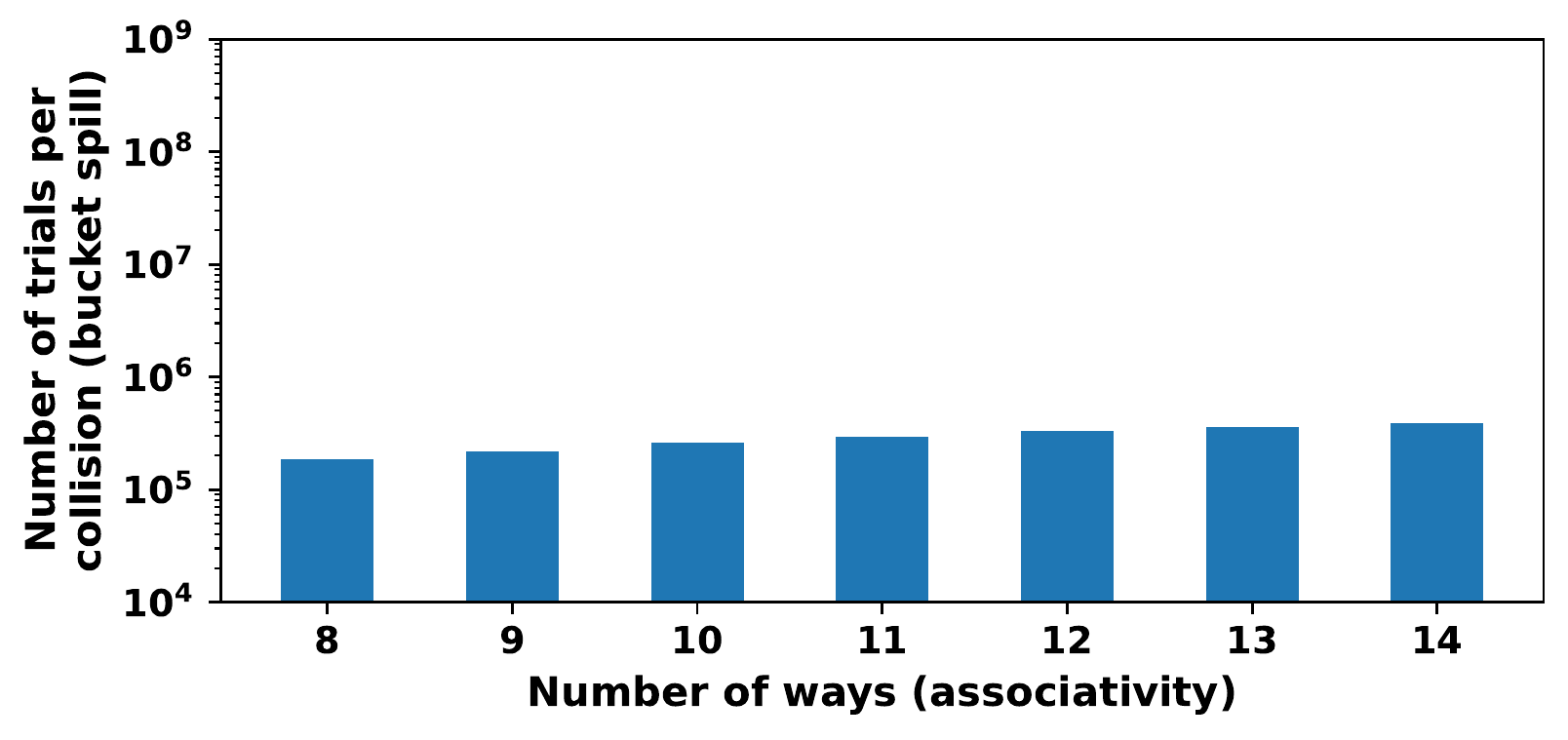}

        \caption{Original Code (with assert failure)}
        \label{fig:figure1}
    \end{subfigure}
    \hfill
    \begin{subfigure}[b]{0.49\textwidth}
        \includegraphics[width=\textwidth]{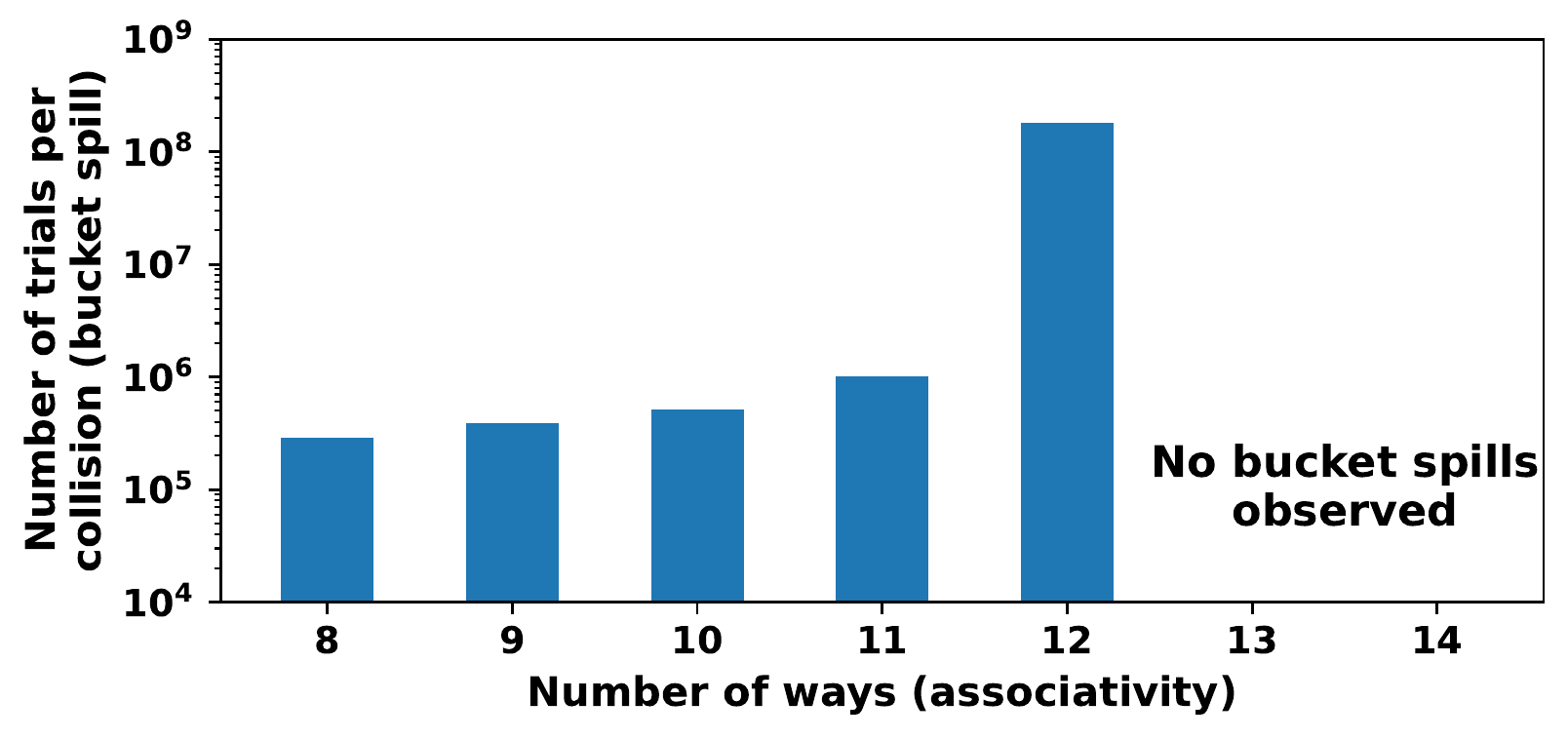}
        \caption{After Bug-Fix}
        \label{fig:figure2}
    \end{subfigure}
    \caption{Number of trials before a collision (bucket spill) -- Figure-6 in HPCA'23 paper. (a) In the original code~\cite{randcachebroken}, the trials before a collision grows linearly with associativity. (b)~After the bug-fix, it is super-exponential; we observe no bucket spills at associativity of 14 (8 $+$ 6 extra ways), the default in Mirage~\cite{mirage}.}
    \label{fig:bnb_fix}
\end{figure}

In \cref{fig:bnb_fix}, we reproduce results from Figure~6 of the HPCA'23 paper. In the original results (which had assert failures) in \cref{fig:figure1}, the number of ball throws (trials) before a collision (bucket-spill or set-conflict) grows linearly. With the bug fix and the correct $remove\_ball()$ implementation, we obtain the results in \cref{fig:figure2} showing a super-exponential trend with increasing associativity. We observe no bucket spills after 1 billion trials with associativity of 14 ways (or 13 ways), which is the default for MIRAGE. 
 
\subsection{Bugs in the Python Cache Simulator}
In our email to the authors of the HPCA'23 paper on February 21, 2023, we also pointed out that their python cache simulator does not implement Global Evictions in MIRAGE, and as a result suffers from the same assert failures (lines in cache exceeding the cache capacity) similar to \cref{fig:bnb_assert_failure}, invalidating their results.  While the authors responded in a couple of days with an updated code implementing global evictions, we identified two additional bugs in their Python cache simulator that further invalidated Figure 7 in the paper. 

\begin{figure}[h]
    \centering
    \includegraphics[width=\textwidth]{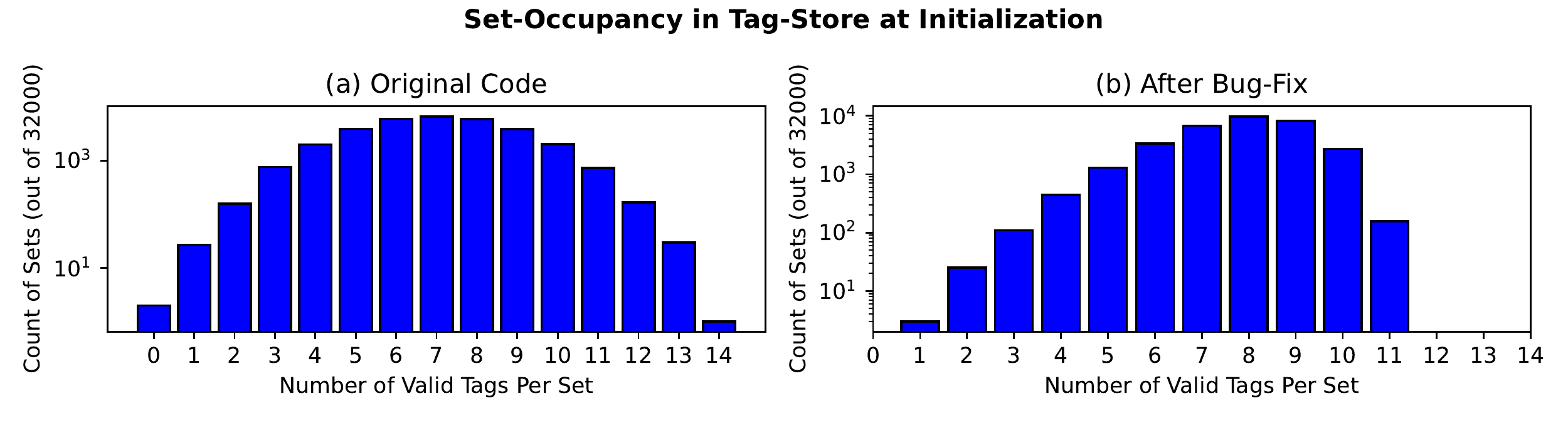}
    \caption{Bug in Initialization of Tags. (a) Original code assumes each tag can be valid with probability=0.5, which results in already full sets at initialization. (b) Bug fix: Cache starts from an invalid state and has K random addresses inserted and results in a more representative set occupancy  in tag-store at initialization.}
    \label{fig:python_bug_init}
\end{figure}

\textbf{Bug-1.} The first bug is that the tag-store is initialized by assigning each tag valid with a probability of 50\%. However, this results in high occupancy levels in a few sets, which is virtually impossible under load balancing. \cref{fig:python_bug_init}(a) shows the set-occupancy levels for different sets at initialization in the original simulator code, and in fact the HPCA'23 starts, in expectation, with a state where more than one set is full at initialization time. In essence, to break MIRAGE, the paper assumes an already broken state! 

The correct initialization methodology for the tag-store is to start with an invalid cache and insert K random addresses through the cache interface including load-balancing and global evictions. 
\cref{fig:python_bug_init}(b) shows the initial set-occupancy levels in the simulations after this bug fix, where the high occupancy buckets are prevented by Mirage.

\textbf{Bug-2.} We identified a second bug in the cipher used for randomizing the cache set indexing in the python cache simulator. MIRAGE requires a uniformly random set-index derivation function for security as noted in the original paper. 
However, we discovered that the cache set indices generated in the HPCA'23 paper's code were not sufficiently uniform random. 
The root-cause was a buggy implementation of the PRESENT cipher used for the cache set indexing in the author's code base. We tested it with known test vectors from the PRESENT paper~\cite{PRESENT} and the HPCA’23 paper’s implementation fails those tests, generating incorrect ciphertexts for known plaintext and keys. We illustrate this in a screenshot in \cref{fig:python_bug_index} in the Appendix.
Replacing this buggy cipher with a standard implementation of AES-128 or using PRINCE-64, which is used in the MIRAGE paper, with random keys, addressed this issue.

\cref{fig:python_bug_index} shows the distributions in the set index for 1 million random addresses with the original code and after the bug fix. 
\cref{fig:python_bug_index}(a) shows considerable skew in distribution of indices across possible values in the original code.  
Replacing this cipher with AES-128 or PRINCE-64, which is used in the MIRAGE paper, with random keys, we were able to produce uniform set-indices across addresses, as shown in \cref{fig:python_bug_index}(b,c).

\begin{figure}[h]
    \centering
    \includegraphics[width=\textwidth]{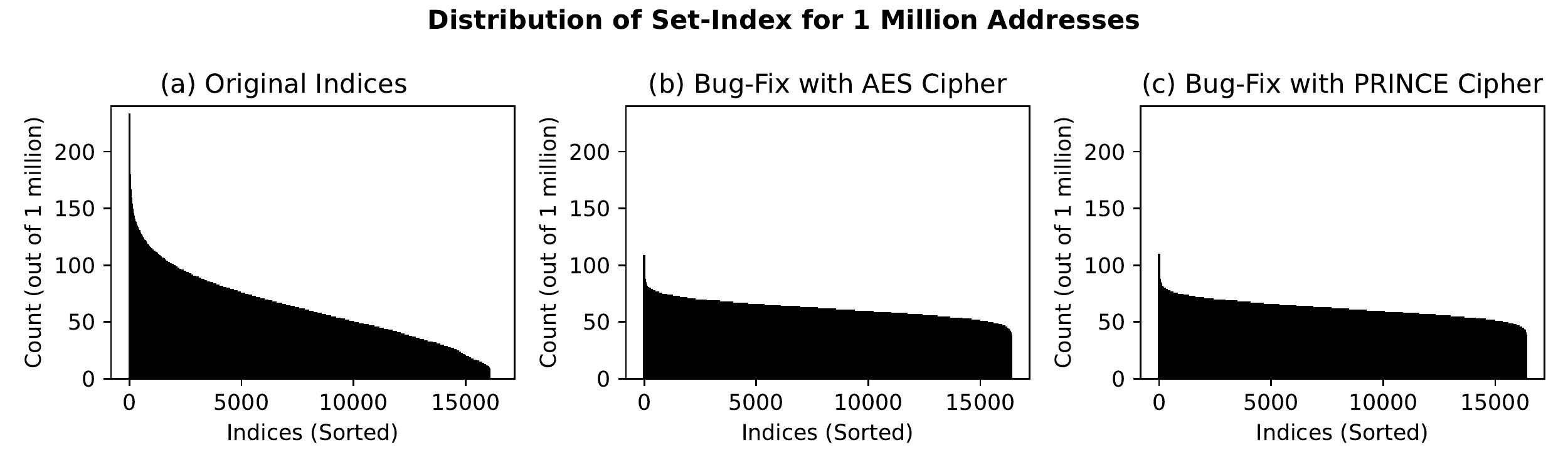}
     \caption{Bug in Set-Index Derivation Causing Non-Uniform Indices. A key requirement for MIRAGE to be secure is a uniformly random cipher. Across 1 million random addresses, (a) the cipher in the original code has a skewed distribution with a std dev of 31 that is not uniformly random. (b,c) Using out-of-the-box AES or PRINCE cipher in the index derivation function results in a uniform distribution, with a std dev of 7.9.}
  
    \label{fig:python_bug_index}
\end{figure}

\textbf{Sanity Check with Analytical Model:}  We can corroborate the results from Fig~\ref{fig:python_bug_index} using a simple analytical model.  If 1 million balls (lines) are thrown in 16,364 buckets (sets), then the average ($\mu$) is equal to 61.  Given this is a Poisson process, we can estimate the standard deviation ($\sigma$) across buckets (sets) to equal the square root of the mean, thus $\sigma = \sqrt{61} = 7.8$. We observe that with AES and PRINCE indeed we get a $\sigma$ of 7.9, which is close to the expected value. For a well-behaved process, we expect values that are 6 standard deviations away from the mean to be highly improbable (about 1 in a billion), so we should not expect to see values above $61+ 6 \cdot 7.9 = 108$, and that is indeed the case with AES and PRINCE.

However, with the author's implementation of PRESENT, we get a $\sigma$ of 31, denoting a non-Poisson process with very high variation across buckets (sets). Indeed, we do see values exceeding 200 with the PRESENT implementation, indicating significant non-uniformity across sets.  This non-uniformity breaks the assumption of MIRAGE that the cache index is based on a randomized function that uniformly routes addresses to sets.

\section{Impact of Bug-Fixes} 

After fixing these bugs -- ensuring a correct global eviction implementation, correct initialization of tag-store, and using a functionally correct cipher (AES/PRINCE) -- in the author's python cache simulator, we observe no set-associative evictions (valid tag evictions) for varying cache sizes. This further invalidates the claims made in the paper that Mirage is broken.

\begin{figure}[htbp]
    \centering
    \begin{subfigure}[b]{0.49\textwidth}
        \includegraphics[width=\textwidth]{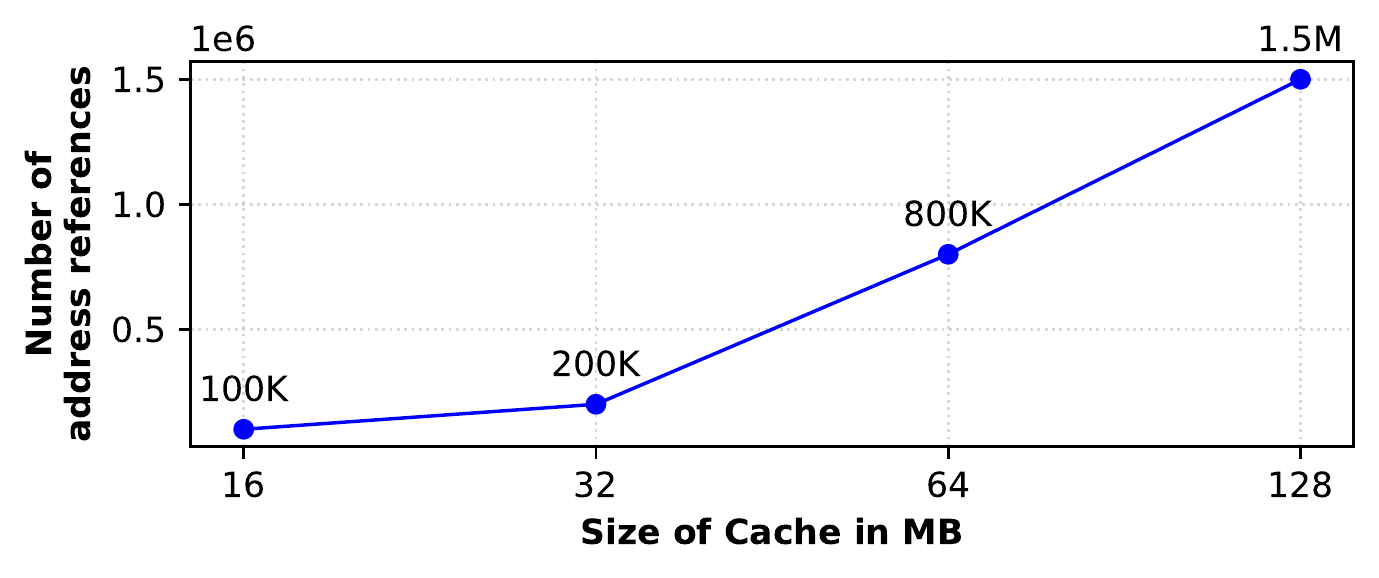}
        \caption{Original Code}
        \label{fig:figure7a}
    \end{subfigure}
    \hfill
    \begin{subfigure}[b]{0.49\textwidth}
        \includegraphics[width=\textwidth]{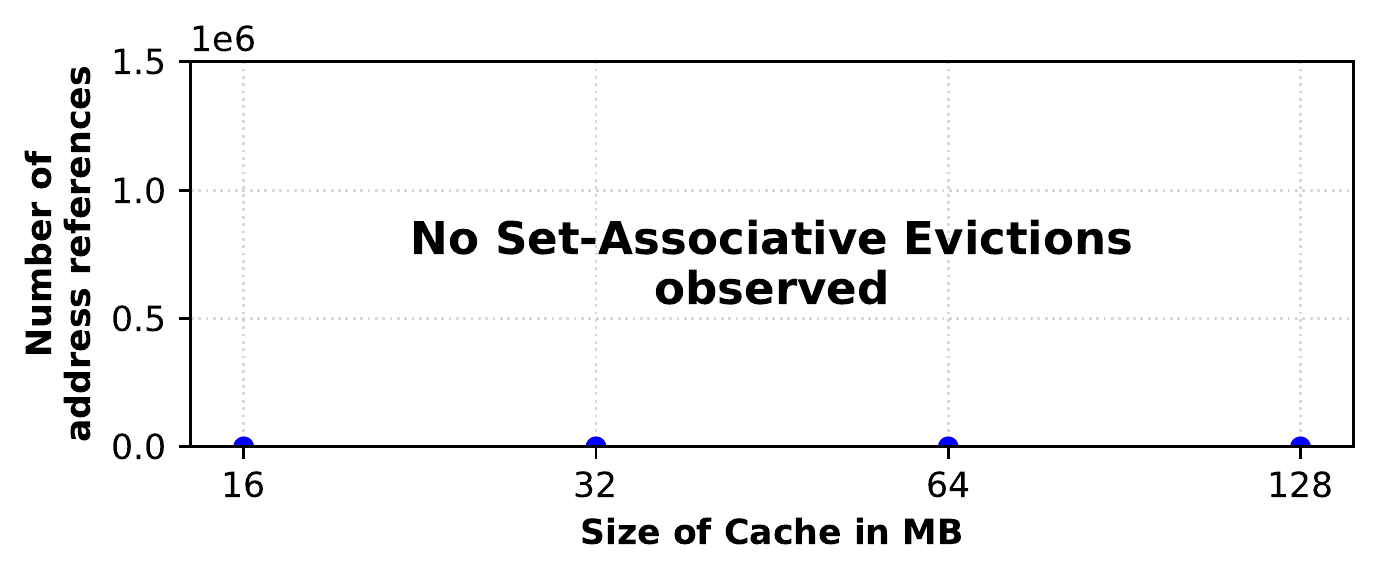}
        \caption{After Bug-Fix}
        \label{fig:figure7b}
    \end{subfigure}
    \caption{Number of cache references before a set-associative eviction (valid tag eviction) -- Figure-7 in HPCA'23 paper. (a) In the original code~\cite{randcachebroken}, the cache references before a set-associative eviction grows linearly with cache size. (b)~After the bug-fix, there are no observed set-associative evictions at a associativity of 8$+$6 ways in MIRAGE.}
    \label{fig:pyc_fix}
\end{figure}

\section{Conclusion}

In conclusion, we provided evidence to support our claim that the MIRAGE cache is not broken. The root-cause of the issues perceived by the HPCA'23 paper were bugs in the  modeling and simulation framework by their authors. We acknowledge and thank the authors of the HPCA'23 for sharing their code publicly. 

\newpage
\appendix
\section{Assert Failure in Bucket and Balls Simulation}
\begin{figure}[h]
    \centering
    \includegraphics[width=\textwidth]{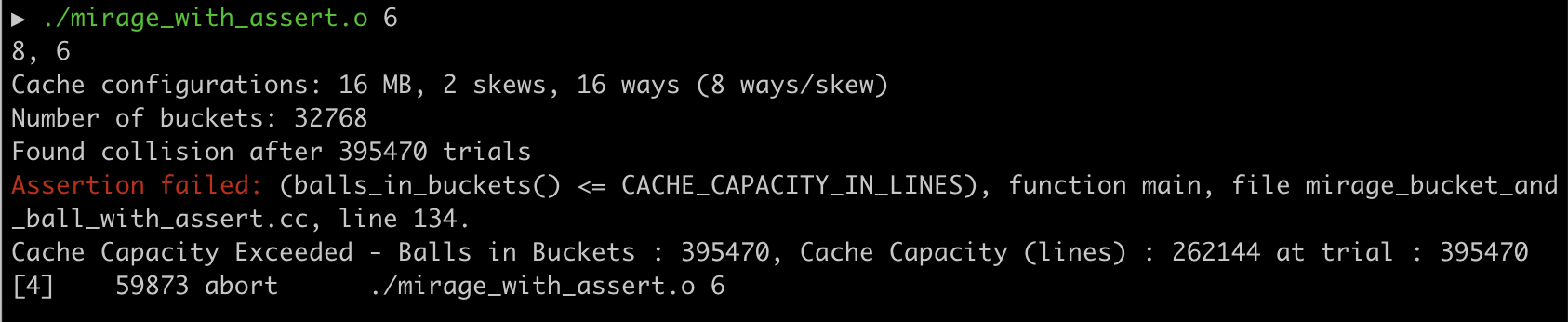}
    \caption{Output of the original Buckets and Balls code with an added assertion. At the time of a collision (bucket spill), the code has installed 385,000 addresses into a cache with a capacity of 256,000 cache lines, which leads to an assertion failure. }
    \label{fig:bnb_assert_failure}
\end{figure}

\section{Bug in Set Index Derivation in Python Cache Simulator}

\begin{figure}[h]
    \centering    \includegraphics[width=\textwidth]{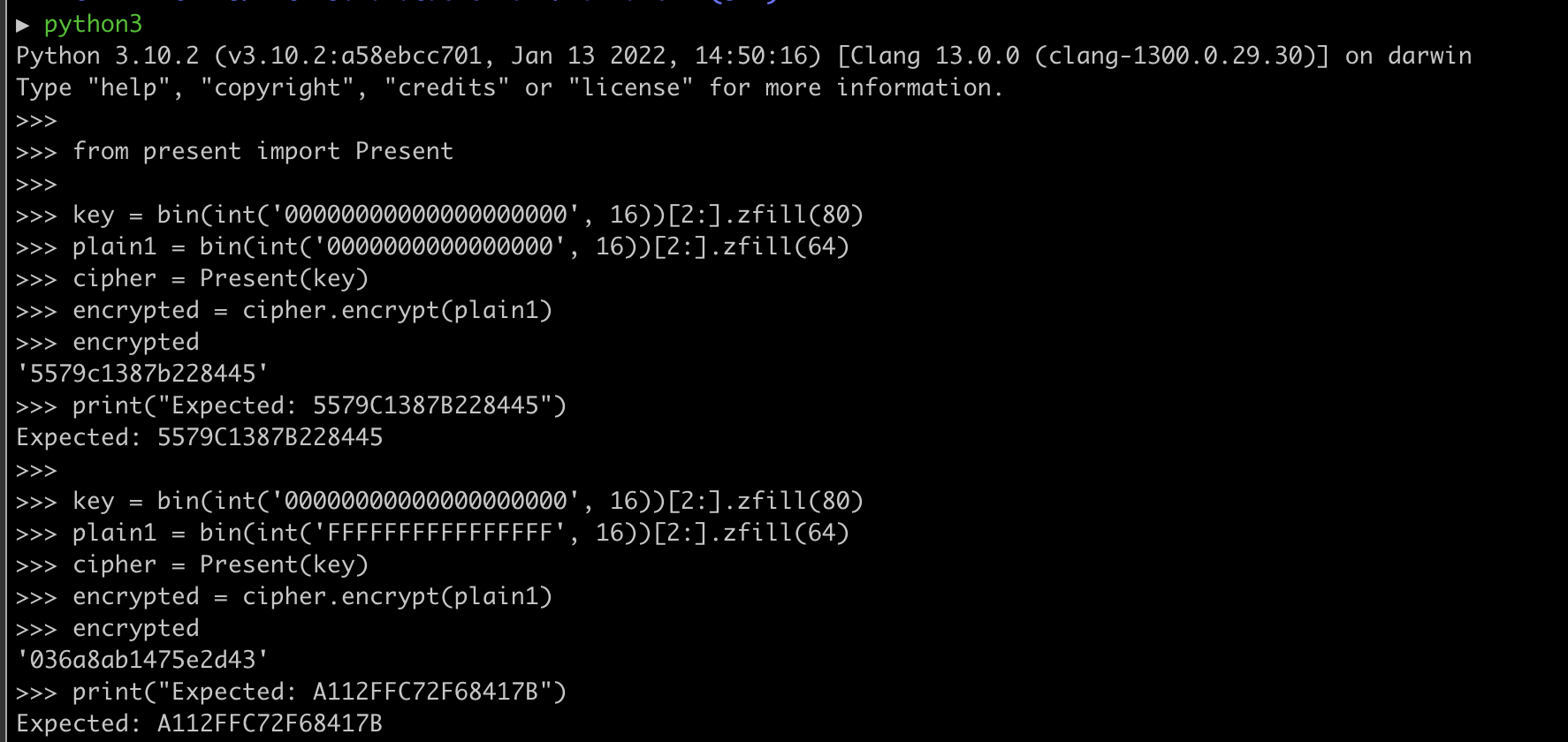}
     \caption{Bugs observed in the PRESENT cipher implementation in the HPCA'23 paper~\cite{randcachebroken}. We tested it with test vectors from the PRESENT paper~\cite{PRESENT} and the HPCA'23 paper's implementation fails those tests. It produces an incorrect cipher-text for key $=$ \texttt{0x00000000000000000000} and plaintext $=$ \texttt{0xFFFFFFFFFFFFFFFF}) -- we expected \texttt{0xA112FFC72F68417B}, but obtained \texttt{0x036A8AB1475E2D43}.} 
  
    \label{fig:python_assert_index}
\end{figure}


\bibliography{ref}
\end{document}